\documentclass[%
reprint,
superscriptaddress,
nofootinbib,
amsmath,amssymb,
aps
]{revtex4-1}
\usepackage{dcolumn}
\usepackage{bm}
\usepackage{epsfig}
\usepackage{amsmath}
\usepackage{subfigure}
\usepackage{amsfonts}
\usepackage{amssymb}
\usepackage{graphicx}
\usepackage{pstricks}
\usepackage{array}
\usepackage{hyperref}
\usepackage{multirow}

\DeclareSymbolFont{AMSa}{U}{msa}{m}{n}
\DeclareSymbolFont{AMSb}{U}{msb}{m}{n}
\let\Box\relax
\DeclareMathSymbol{\Box}{\mathord}{AMSa}{"03}

\def\IR{{\mathbb R}}

\newcommand{\be}{\begin{equation}}
\newcommand{\ee}{\end{equation}}
\newcommand{\bea}{\begin{eqnarray}}
\newcommand{\eea}{\end{eqnarray}}
\newcommand{\f}{\frac}
\newcommand{\Sum}{\displaystyle\sum\limits}
\newcommand{\ds}{\displaystyle}

\newcolumntype{L}[1]{>{\raggedright\let\newline\\\arraybackslash\hspace{0pt}}m{#1}}
\newcolumntype{C}[1]{>{\centering\let\newline\\\arraybackslash\hspace{0pt}}m{#1}}
\newcolumntype{R}[1]{>{\raggedleft\let\newline\\\arraybackslash\hspace{0pt}}m{#1}}

\begin{document}

\preprint{UTTG-26-14, TCC-026-14, MIT-CTP/4615}

\title{Analyzing Multi-Field Tunneling With Exact Bounce Solutions}

\author{Aditya Aravind}
\email{aditya@physics.utexas.edu}
\affiliation{Department of Physics and Texas Cosmology Center\\ The University of Texas at Austin, TX 78712, USA}
\author{Brandon S. DiNunno}
\email{bsd86@physics.utexas.edu}
\affiliation{Department of Physics and Texas Cosmology Center\\ The University of Texas at Austin, TX 78712, USA}
\author{Dustin Lorshbough}
\email{lorsh@utexas.edu}
\affiliation{Department of Physics and Texas Cosmology Center\\ The University of Texas at Austin, TX 78712, USA}
\author{Sonia Paban}
\email{paban@zippy.ph.utexas.edu}
\affiliation{Department of Physics and Texas Cosmology Center\\ The University of Texas at Austin, TX 78712, USA}
\affiliation{Center for Theoretical Physics and Department of Physics, \\ Massachusetts Institute of Technology, Cambridge, MA 02139, USA }

\date{\today}

\begin{abstract}
We study multi-field tunneling using exact solutions for additive potentials. We introduce a binomial potential with non-integer powers that could be considered a generalization of the $4D$ Fubini instanton potential. Using scaling arguments, we show that for multi-field potentials taller and wider barriers may still lead to a smaller bounce action.
\end{abstract}

\pacs{Valid PACS appear here}


\maketitle

\section{Introduction}
The string theory landscape motivates the study of multi-field potentials with a large number of metastable vacua \cite{Sarangi:2007jb,Johnson:2008kc,HenryTye:2008xu,Tye:2009rb,Balasubramanian:2010kg,Czech:2011aa,Silverstein:2008sg,Bousso:2000xa,Feng:2000if,Ahlqvist:2010ki,Duncan:2014oja}. Since our universe may have occupied a metastable vacuum in the past or may do so today, it is of cosmological interest to study tunneling out of metastable vacua  \cite{Linde:1998gs,Freivogel:2005vv,Guth:2007ng,Freivogel:2011eg,Yamauchi:2011qq,Park:2011ty,Sugimura:2013cra,Bousso:2013uia,Bousso:2014jca}. As the tunneling rate depends on the action of the bounce solution \cite{Coleman:1977py}, computing the bounce action for various tunneling scenarios is of interest for studying phase transitions in the early universe.

Calculating the exact bounce action is difficult for general multi-field potentials. Sarid \cite{Sarid:1998sn} and later Greene et al. \cite{Greene:2013ida} discussed approaches towards estimating bounce action for quartic potentials using semi-numerical methods motivated by analytical arguments. Greene et al. \cite{Greene:2013ida} further concluded that for quartic potentials the bounce action decreases as a power law in the number of fields, agreeing with the lower bound calculated later in \cite{Aravind:2014aza}. 

For multi-field potentials, part of the difficulty in calculating the bounce action arises from the difficulty in determining the appropriate tunneling trajectory. The problem could therefore be made more manageable by splitting it into two smaller problems - determining the field-space trajectory of least action from the potential, and determining the bounce action along the least action trajectory. In this paper, we discuss the second problem some detail.

Studying the dependence of the bounce action on the potential profile along the path would be facilitated by studying potentials with exact analytical bounce solutions. We present such a potential in the form of a binomial with non-integer powers which can be considered a generalization of the Fubini instanton (see, for example, \cite{Loran:2006sf}); unlike the standard Fubini instanton, this potential has a barrier through which the field tunnels out. We also discuss a potential which has been previously known \cite{FerrazdeCamargo:1982sk}. Through these examples of single field potentials with exact solutions, we study how barrier features are related to the bounce action. We see that a taller and wider barrier could lead to a smaller bounce action, contrary to what is expected from non-relativistic quantum mechanics. 

We then introduce a scaling argument that helps identify how the bounce action scales with barrier parameters for a general single field potential. We show how this is consistent with the results we have for the exactly solvable potentials. By extending this argument to additive multi-field potentials, we discuss the accuracy of the approximation scheme used by Greene et al. \cite{Greene:2013ida}. 

The rest of this paper is organized as follows. In section \ref{sec:basics}, we briefly review tunneling in field theory. In section \ref{sec:exactsol}, we discuss special potentials for which an analytic bounce solution is available. In section \ref{sec:scaling}, we introduce a scaling argument for bounce action of single field potentials and discuss its implications. In section \ref{sec:multipot}, we apply this argument to gain insights on tunneling in multi-field potentials.  In section \ref{sec:conclusion}, we conclude.

\section{Review of Tunneling} \label{sec:basics}

In this section, we briefly review tunneling in field theory in 4-dimensional Euclidean space, following the approach by Coleman \cite{Coleman:1977py}. For the rest of this paper, we always assume that the field(s) has a metastable vacuum at the field-space origin $\vec{\phi} = \vec{0}$, with potential $V(\vec{0})=0$. We assume that this vacuum is surrounded on all sides by a barrier with $V>0$. For tunneling to happen, there must exist regions beyond the barrier with $V<0$ into which the field can tunnel. 

For the case of a single field $\phi$, it was proven in \cite{Coleman:1977th} that tunneling proceeds through the formation of an $O(4)$ symmetric bubble in Euclidean space (more specifically, it was shown that for a wide class of potentials, the action is minimized by $O(4)$ symmetric configurations). Therefore, the field value everywhere in space can be expressed as a function of the radius $r$ measured from the center of the bubble. The field profile $\phi(r)$ obeys the following equation of motion obtained from assuming $O(4)$ symmetry \cite{Coleman:1977py}
\be \label{eq:eom}
\f{d^2}{dr^2}\phi + \f{3}{r}\f{d}{dr}\phi = \f{\partial V}{\partial \phi} \,, \quad \dot{\phi}(0)=0 \, , \quad \phi(\infty)=0 \, .
\ee

The solution to this equation $\bar{\phi}(r)$ (also called the bounce) corresponds to the position of a classical particle moving (in field space) in the inverted potential $\ds{-V(\phi)}$ subject to time-dependent friction. The initial conditions impose that the particle starts at rest from the point where the field tunnels out ($\ds{\phi(0)=\phi_0}$) and ends at rest at the false vacuum ($\phi(\infty) = 0$). The tunneling rate (per unit volume) is given by $\ds{\Gamma /V \sim A \, e^{-S/\hbar }}$. Here $S$ is the bounce action, which can be written as
\bea \label{eq:bounce}
S &=& 2 \, \pi^2 \int_0^\infty dr \hspace*{0.5mm} r^3 \left[\f{1}{2}\left(\f{d\bar{\phi}}{dr}\right)^2 + V(\bar{\phi}) \right] \cr \cr
 &=& \f{\pi^2}{2} \int_0^\infty dr \hspace*{0.5mm} r^3 \left(\f{d\bar{\phi}}{dr}\right)^2 \, .
\eea
Here, the last equality follows from Derrick's theorem \cite{Weinberg:2012pjx}.

For multi-field potentials, we are not aware of a proof for the $O(4)$ symmetry of the tunneling solution. However, $O(4)$ symmetry is generally assumed \cite{Wainwright:2011kj,Greene:2013ida,Aravind:2014aza} (though not always \cite{Balasubramanian:2010kg,Czech:2011aa,Masoumi:2012yy}), and we do the same here. Under this assumption, the story proceeds in a manner analogous to the single field case. Each field coordinate $\phi_i$ obeys the equation 
\be \label{eq:eom2}
\f{d^2\phi_i}{dr^2} + \f{3}{r}\f{d\phi_i}{dr} = \f{\partial V}{\partial \phi_i} \,, \quad \dot{\phi}_i(0)=0 \,, \quad \phi_i(\infty)=0 \, .
\ee

The solution to this set of equations is analogous to the position vector $\vec{\phi}$ of a particle moving subject to friction in an inverted multi-dimensional potential $-V(\vec{\phi})$ along some specific trajectory. Since the position, velocity and acceleration are multi-component vectors, we can reorganize these equations into a more intuitive form by separating the components parallel to and perpendicular to the trajectory of the particle (see, for example, \cite{Wainwright:2011kj}). We begin by parametrizing points on the trajectory in terms of the field-space distance from the false vacuum measured \emph{along the trajectory}\footnote{Note that $\phi(r)$ is the arc-length along the trajectory, and \emph{not} the radial distance from the field-space origin. At any point on the trajectory $\ds{d\phi^2 = \Sum_{i=1}^N d\phi_i^2}$, but in general $\ds{\phi^2(r) \neq \Sum_{i=1}^N \phi_i^2(r)}$.}, $\phi(r)$. In terms of this variable, the equations (\ref{eq:eom2}) can be re-written as 
\bea \label{eq:eom3}
\f{d^2\phi}{dr^2} + \f{3}{r}\f{d\phi}{dr} &=& \f{\partial}{\partial \phi}V(\vec{\phi}) \, , \cr \cr \f{d^2 \vec{\phi}}{d\phi^2}\left(\f{d\phi}{dr}\right)^2 &=& \nabla_\perp V(\vec{\phi})\, .
\eea
Here, $\f{\partial}{\partial \phi}V(\vec{\phi})$ and $\nabla_\perp V(\vec{\phi})$ refer to the tangential and perpendicular components of the gradient of the potential respectively. The first equation is similar to the single field equation of motion (\ref{eq:eom}), while the second equation causes the bounce trajectory to curve (in field-space) when the potential slopes in the transverse directions. If the trajectory is known, the multi-field problem can be treated effectively as a single field problem with a field $\phi$ subject to a potential $V(\phi)$ (the ``potential profile" on the trajectory), with an action identical in form to (\ref{eq:bounce}). 

If there are multiple solutions to (\ref{eq:eom3}), the one with the lowest action (the bounce) typically dominates tunneling. For a general potential profile, it is possible to get a rough estimate/underestimate of the action \cite{Greene:2013ida,Aravind:2014aza} after truncating the integral\footnote{The truncated integral $\ds{\int_0^{\phi_\Sigma} d\phi \sqrt{2V(\phi)}}$ is commonly referred to as ``surface tension".} in (\ref{eq:bounce})
\bea \label{eq:approx}
S & \gtrsim &  \, r_\Sigma^3 \, \f{\pi^2}{2} \int_{r_\Sigma}^\infty dr \left(\f{d\phi}{dr}\right)^2 \, \cr \cr
& \gtrsim &   r_\Sigma^3 \, \f{\pi^2}{2} \int_0^{\phi_\Sigma} d\phi \sqrt{2 \, V(\phi)} \, \cr \cr & \approx & r_\Sigma^3 \, \pi^2 \int_0^{\phi_S} d\phi \sqrt{2 \,V(\phi)} \, .
\eea
Here, $r_\Sigma$ and $\phi_\Sigma$ refer to the values of $r$ and $\phi$ at the point on the trajectory where the potential is equal to its false-vacuum value (if there are multiple such points, we take the one closest to the false vacuum on the trajectory). Similarly, $\phi_S$ refers to the value of $\phi$ at the local maximum of the potential profile closest to the false vacuum. 

In \cite{Greene:2013ida}, the authors write the pre-factor multiplying the surface tension integral in terms of the bounce radius $r$. Since bounce radius is not a well-defined quantity in general (except in the case of a thin-wall bubble), we shall define it to be $r_\Sigma$. As a simplifying assumption, they also argued that tunneling occurs through a nearby saddle point which presents the smallest barrier. When that happens, our definition of $\phi_S$ would correspond to the location of this saddle point. However, we note that the bounce trajectory in general does not have to pass through a saddle point \cite{Konstandin:2006nd}.

\section{Single Field Potentials With Exact Bounce Solutions} 
\label{sec:exactsol}

In this section, we discuss two single field potentials with an exact analytical solution to (\ref{eq:eom}). 

\subsection{Binomial Potential With Non-Integer Powers}

The first example we discuss can be considered a generalization of the Fubini instanton to non-integer powers. The potential is
\bea \label{eq:bino}
V(\phi) = \frac{4 \, u \, n^2(n-1)}{2 \, n+1}\phi^{(2n+1)/n} - 2 \, u \, v \, n^2 \phi^{(2n+2)/n} , \cr 
\eea
where $\ds{\left\lbrace u,v,n \right\rbrace  \in \IR}$ with $\ds{u>0}$, $\ds{v>0}$ and $\ds{n>1}$. 

The exact bounce solution and bounce action for this potential are given by
\bea \label{eq:bino2}
\bar{\phi}(r) &=& \frac{1}{(u \, r^2 + v)^n} \, , \cr \cr
S \left[\bar{\phi} \right] &=&  \f{n \, \pi^2}{(4 \, n^2-1)}\f{1}{u \, v^{2n-1}} \, . 
\eea

Unlike the standard Fubini case, this potential has a minimum at $\phi=0$ followed by a barrier for small values of $\phi$ and a runoff (to $-\infty$) for large values of $\phi$. In general, this potential represents a case of thick-wall tunneling. 

\begin{figure}[ht]

\includegraphics[width=0.95\linewidth]{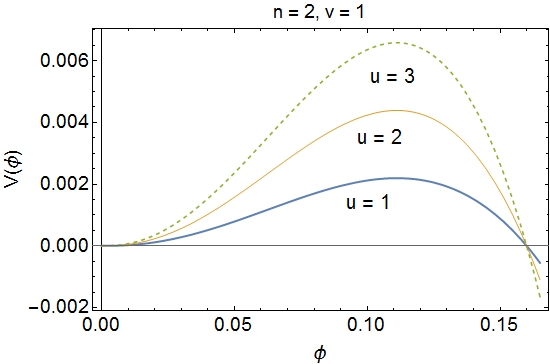} \vspace*{3pt}
\begin{pspicture}(0,-0.1)(1,0.1)
\end{pspicture} 

\includegraphics[width=0.95\linewidth]{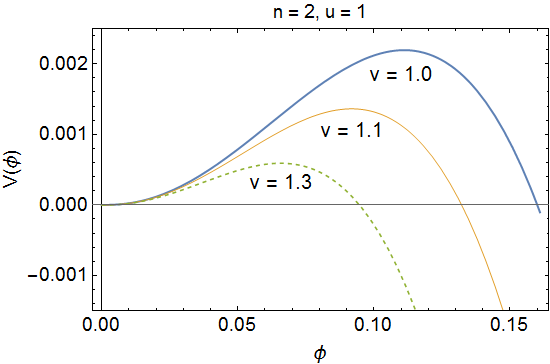} \vspace*{3pt}
\begin{pspicture}(0,-0.1)(1,0.1)
\end{pspicture} 

\includegraphics[width=0.95\linewidth]{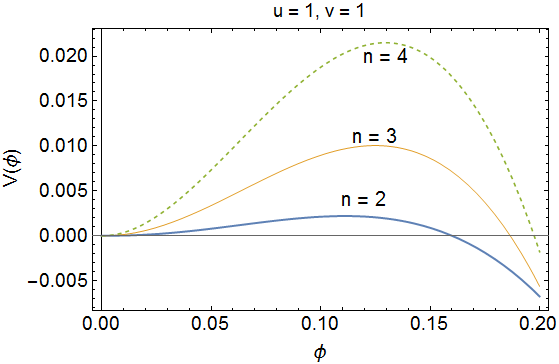}
\caption{\label{fig:bino}Profile of the Binomial potential (\ref{eq:bino}) for different values of $u$ (top), $v$ (center) and $n$ (bottom). Note that $u$ and $v$ rescale the potential while $n$ changes its shape.}
\end{figure}

For physical theories, we might worry about the runoff and also about the behavior of the potential for $\phi<0$. The former could be addressed by adding terms that avoid runoff for large values of $\phi$, and the latter by replacing $\phi$ with $|\phi|$. The bounce itself cares only about the potential profile along the tunneling trajectory (between $\phi=0$ and $\phi = 1/v^n$), not beyond it. 

We notice that for $\ds{0<n\leq 1}$ the potential is well defined, but it does not have a barrier. Therefore the solution does not involve tunneling. For $\ds{n=1}$ we recover the Fubini case. For larger integer values of $n$, the potential involves fractional powers and bears some resemblance to potentials encountered in the string theory landscape \cite{Silverstein:2008sg}. It is worth investigating whether potentials with exactly this form - a binomial with two non-integer powers between 2 and 4 - appear somewhere in physically relevant situations.

\begin{figure*} 
\centering
\includegraphics[width=0.8 \textwidth]{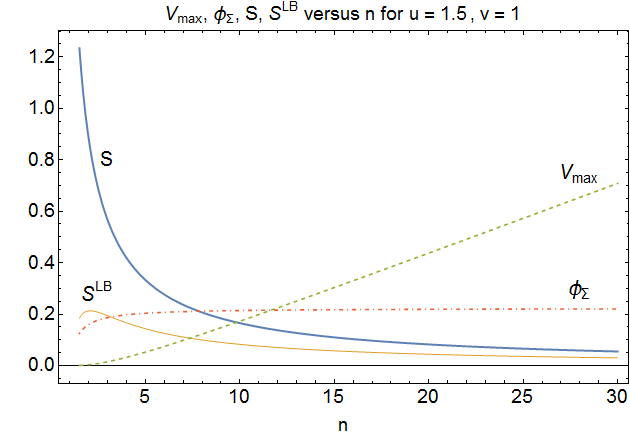}
\caption{\label{fig:nscaling}Scaling of various quantities with $n$ for the binomial potential (\ref{eq:bino}), with $u=1.5$, $v=1$. The four lines represent action $S$ (thick, blue), approximation $S^{LB}$ (thin, orange), height of the potential peak $V_{max}$ (dashed, green) and barrier width $\phi_{\Sigma}$ (dot-dashed, red). Note that $S^{LB}$ has been scaled up by a factor of 5 for ease of comparison.}
\end{figure*}

We note that increasing the value of the parameter $u$ corresponds to scaling up the potential, while increasing $v$ corresponds to scaling down the potential \emph{as well as} making the barrier narrower as seen in Fig. \ref{fig:bino}. Both of these changes tend to bring down the bounce action. If we fix $u$ and $v$ and vary the parameter $n$, the dependence is more complicated. To take a special case, if we fix $v=1$, we see that as we increase $n$, the barrier width and the barrier height increases (the width asymptotes to a constant) while the bounce action decreases, as shown in Fig. \ref{fig:nscaling}. Hence, a higher barrier can result in a lower action, as known from scaling relations \cite{Weinberg:2012pjx}. The dependence of bounce action on scaling of barrier parameters will be studied in greater generality in section \ref{sec:scaling}. 

\subsection{Logarithmic Potential}

We shall now discuss a potential with exact bounce solution which has been known \cite{FerrazdeCamargo:1982sk}. 

\bea \label{eq:logpot}
V(\phi) &=& \f{1}{2}m^2 \phi^2 \left[ 1 - \ln \left(\f{\phi^2}{w^2}\right) \right] \, , \quad \left\lbrace m, w \right\rbrace \in \IR \, . \cr & & 
\eea
The bounce solution is given by
\bea \label{eq:logpot2}
\bar{\phi}(r) &=& w \exp \left[ - \f{1}{2}m^2 r^2 + 2 \right] \, , \cr \cr
S &=& \f{\pi^2 e^4}{2}\f{w^2}{m^2} \, .
\eea
Similar to the binomial potential, this potential has a barrier near the false vacuum followed by a run-off for large values of $\phi$. The $\log$ function has a singularity at $\phi=0$, but the potential has a well defined limit at that point. Scaling the parameters $w$ and $m$ correspond to scaling the field-space width of the barrier or scaling the height of the potential (with corresponding changes to the radius-scale $r_\Sigma$). How the bounce action depends on these parameters is also clear; the bounce action decreases with the barrier height and increases with the barrier width. We shall generalize this discussion in the next section.

\section{Scaling of Bounce Action With Potential Profile}\label{sec:scaling}

\subsection{Scaling Argument}

In this section we shall discuss how the scaling of the potential profile along the bounce trajectory (which changes the height and width of the barrier) affects the bounce action. Since this is effectively a single field problem, we shall call the field variable $\phi$, corresponding to either the single field coordinate (for the single field potential) or the arc length along a multi-field trajectory (for the multi-field potential). The potential profile along the trajectory includes the barrier region ($\ds{V>0}$) and also the region beyond it ($\ds{V<0}$).

Let us start with a tunneling potential $V(\phi)$ with a bounce solution $\ds{\bar{\phi}(r)}$ (not necessarily analytic). In order to parametrize scaling, we introduce a variable $\ds{g>0}$, with $\ds{g=1}$ corresponding to the original potential. On changing the value of $g$, let the potential $V$ (barrier height) and the length-scale in field-space $\phi$ (barrier width) scale as powers of $g$, $\ds{g^a}$ and $\ds{g^b}$ respectively. We denote the rescaled potential and bounce solution as  $\ds{V_g}$ and $\ds{\bar{\phi}_g}$ respectively ($\ds{V_1 \equiv V}$ and $\ds{\bar{\phi}_1 \equiv \bar{\phi}}$). We observe that for the scaling to be consistent, the typical radius $r$ must also scale. Collectively, the scaling relations are as follows
\bea \label{eq:scaling}
r_g \hspace*{2mm} &\equiv & \hspace*{2mm} g^c \hspace*{1mm} r \, , \cr
\phi_g \hspace*{2mm} &\equiv & \hspace*{2mm} g^b \hspace*{1mm} \phi \, , \cr 
\bar{\phi}_g(r_g) \hspace*{2mm} &=& \hspace*{2mm} g^b \hspace*{1mm} \bar{\phi}(r) \, , \cr 
V_g(\phi_g) \hspace*{2mm} &=& \hspace*{2mm} g^a \hspace*{1mm} V(\phi) \, . 
\eea

In order to ensure that $\bar{\phi}_g$ satisfies the equation of motion (\ref{eq:eom}), we must have an additional constraint
\be \label{eq:abc}
2 \, c \hspace*{2mm} = \hspace*{2mm} 2 \, b - a \, .
\ee

This is consistent with the scaling argument presented in \cite{Scrucca:2012cs} for the single-field case. From the scaling relations (\ref{eq:scaling}) and (\ref{eq:abc}) and from (\ref{eq:bounce}), the bounce action scaling is obtained
\bea \label{eq:Sscaling}
S_g = g^{4b-a}S \, .
\eea

We may also check the scaling of the approximation/lower bound (\ref{eq:approx}). Naming this quantity as $\ds{S^{LB}}$, we observe that it scales as $\ds{S_g^{LB} = g^{4b-a}S^{LB}}$, which is the same as the bounce action. Therefore, \emph{any change} in the potential profile and the bounce solution, provided it can be reduced to a scaling of the height of the potential profile and/or a scaling of the field space length scale will maintain the level of accuracy of the approximation. Changes to the shape of the potential profile (which cannot be reduced to some form of scaling) can, however, affect the accuracy of the approximation (\ref{eq:approx}).

\subsection{Application to Exact Solutions}

Let us now apply the scaling argument for the potentials with exact solutions discussed in section \ref{sec:exactsol}. For the logarithmic potential (\ref{eq:logpot}), if the parameters scale as $w_g = g^\gamma w$ and $m_g = g^\delta m$, $\phi$ and $r$ can be rescaled as \begin{eqnarray*}
r_g &\equiv &  g^{-\delta} r \, , \cr
\phi_g &\equiv &  g^{\gamma} \phi \, ,
\end{eqnarray*}
to obtain
\bea 
\bar{\phi}_g(r_g) &=& g^\gamma w \exp \left[ - \f{1}{2}g^{2\delta}m^2 r_g^2 + 2 \right] = g^\gamma \bar{\phi}(r) \, , \cr \cr
V_g(\phi_g) &=& \f{1}{2}g^{2\delta} m^2 \phi_g^2 \left( 1 - \ln \f{\phi_g^2}{g^{2\gamma}w^2} \right) = g^{2(\gamma + \delta)}V(\phi) \, . \cr & &
\eea

\begin{table*}[t] 
\begin{centering}
{
\renewcommand{\arraystretch}{1.4}
\begin{tabular}{ | C{2cm} | C{2.5cm} | C{2.2cm} | C{2.2cm} | C{2.2cm} | C{3cm} |}
\hline
\multirow{2}{*}{Potential} & \multirow{2}{*}{Transformation} & $V$-scaling & $\phi$-scaling  & $r$-scaling &  \multirow{2}{*}{Action-scaling} \\ \cline{3-5}
& & $a$ & $b$ & $c$ & \\
\hline
\multirow{2}{*}{Binomial} & $\ds{u_g = g^\alpha u}$ & \multirow{2}{*}{$\ds{\alpha -(2n+1) \beta}$} & \multirow{2}{*}{$-n \beta$} & \multirow{2}{*}{$\ds{\f{1}{2}(\beta - \alpha)}$} & \multirow{2}{*}{$\ds{S_g = g^{-\alpha - (2n-1)\beta}S}$} \\ 
 & $ \ds{v_g  = g^\beta \, v}$ & & & & \\
\hline
\multirow{2}{*}{Logarithmic} & $\ds{w_g = g^\gamma w} $ & \multirow{2}{*}{$\ds{2(\gamma+\delta)}$} & \multirow{2}{*}{$\gamma$} & \multirow{2}{*}{$-\delta$} & \multirow{2}{*}{$\ds{S_g = g^{2(\gamma-\delta)}S}$} \\ 
 & $\ds{m_g = g^\delta m}$ & & & & \\
\hline
\end{tabular}
}
\caption{\label{tab:scaling}Summary of scaling relations for the binomial (\ref{eq:bino}) and logarithmic (\ref{eq:logpot}) potentials.}
\end{centering}
\end{table*}

This gives us all the scaling exponents in (\ref{eq:scaling}) and (\ref{eq:Sscaling}). This same approach can be followed for the binomial potential. The results are summarized in Table \ref{tab:scaling}.

We note that the $g$-scaling of the action in Table \ref{tab:scaling} agrees with the exact expressions (\ref{eq:bino2}) and (\ref{eq:logpot2}) if we plug in the scaling of parameters $u$, $v$, $w$ and $m$. This is necessary to ensure the consistency of the scaling approach. 

\subsection{Implications of the Scaling Argument}

We shall now discuss the insights gained from the scaling argument regarding the relation between the barrier parameters and the bounce action. We begin by noting that the bounce action scales as $g^{4b-a}$. This means that in general, taller ($a>0$) and narrower ($b<0$) barriers lead to a smaller bounce action. In fact, it is possible to make the barrier larger in all respects (increase in both height and width) and still reduce the bounce action provided the barrier height increases fast enough ($a>4b$) to compensate for the increase in width. The reason for this becomes clear by recalling the estimate (\ref{eq:approx}): the action decreases because the decrease in $r_\Sigma^3$ is faster than the increase in the surface tension integral. 

We note that this is different from the case of non-relativistic quantum mechanics, where barriers with larger surface tension always lead to a larger action due to the absence of the $r^3$ factor\footnote{The surface tension in both cases scales as $\ds{\sigma_g = g^{(a+2b)/2}\sigma}$.}. Therefore, in the case of multi-field potentials, we cannot assume tunneling happens in the direction of the ``smallest'' barrier. In section \ref{sec:multipot}, using additive potentials, we explicitly show situations where the tunneling trajectory lies in the direction of a larger barrier. 

While the scaling argument captures the dependence of action on two barrier parameters (height and width), the actual diversity in the types of potential profiles is far greater than what can be described using only two parameters. Changing the value of $n$ in the binomial potential (\ref{eq:bino}) provides one such example that leads to a non-scaling change in shape of the profile, as seen in Fig. \ref{fig:bino}. The dependence of barrier parameters and bounce action on $n$ can be seen in Fig. \ref{fig:nscaling}. It can be clearly seen that for small values of $n$, the exact bounce action $S$ and the approximation $\ds{S^{LB}}$ scale differently. We note that for $\ds{n \gg 1}$, changing $n$ reduces to a scaling of the potential, which explains why these curves scale the same way for large values of $n$.

\section{Application to Multi-Field Potentials} \label{sec:multipot}

\subsection{Additive Potentials}

One of the simplest ways of going from a single field potential to an $N$-field potential is by defining an additive potential
\be \label{eq:Vadd}
V\left(\vec{\phi} \right) = V\left(\phi_1, \phi_2, ... \phi_N \right) = \Sum_{i=1}^N V_i\left(\phi_i \right) \, .
\ee
Owing to the fact that the different field coordinates $\phi_i$ behave as independent, uncoupled fields, the $N$ equations (\ref{eq:eom2}) completely decouple to give $N$ independent single field equations of motion of the type (\ref{eq:eom}). 

For simplicity, let us assume that each of the $\ds{V_i}$'s has a non-trivial tunneling solution $\ds{\bar{\phi}_i(r)}$ apart from the trivial one ($\ds{\phi_i(r)=0}$). This means that the there are $\ds{2^N-1}$ solutions for the $N$-field bounce, corresponding to each field picking either the trivial or non-trivial solution\footnote{We discount the solution $\phi_i(r)=0 \; \forall \, i$ which does not involve tunneling out of the false vacuum.}.

For additive potentials, the action for $N$-fields is obtained by adding the single field action for each of the $N$ fields (\ref{eq:bounce}) which are completely independent of each-other 
\be
S = \Sum_{i=1}^N \int_0^\infty dr \, r^3\left[\f{1}{2}\left(\f{d\phi_i}{dr}\right)^2 +V_i\left( \phi_i(r) \right) \right] = \Sum_{i=1}^N  S_i \, .
\ee

Since each of the $S_i$'s can be either positive or $0$, the lowest action is provided by the solution where one of the $S_i$'s takes the smallest non-trivial value and all others are $0$; this is the bounce action. 

Effectively, tunneling happens along (or is dominated by) a field axis which corresponds to the $V_i$ that minimizes the bounce action among the $N$ choices available. This does \emph{not} have to correspond to the axis with the smallest barrier. For example, let us consider an $N$-field potential where each of the $V_i$'s is a binomial potential (\ref{eq:bino}) with the same value of $n$ but with different $u$ and $v$ (which scale as powers of $g$). If the exponents in Table \ref{tab:scaling} are such that $\ds{\left\lbrace \alpha>0, \hspace*{0.1mm} \beta < 0 \right\rbrace}$ and $\ds{\alpha > (2n-1)|\beta|}$, the axis corresponding to highest value of $g$ would dominate tunneling. This is not the direction with the smallest barrier as all of the other field axes have shorter and narrower barriers.

\subsection{N-Dependence}

We may study the dependence of the bounce action on the number of fields $N$ to see how $N$ affects vacuum metastability. The answer will depend on what class of additive potentials we consider, i.e., what restrictions we put on them. Let us consider building an $N$-field additive potential of the form (\ref{eq:Vadd}), starting from individual single field components $V_i(\phi_i)$. For simplicity, we assume the $V_i$'s are all potentials of the same type (for example, potentials such as (\ref{eq:bino}) or (\ref{eq:logpot})) differing only in the choice of parameters (such as $u$, $v$, $m$ or $w$), which could all be selected from random distributions. By this choice, we are requiring that the potential profile along each of the axes will have the same shape (but can have different scaling). 

We shall try to enforce some measure of $N$-independence in the potential that we construct by requiring that for any $N$, the typical values of the potential (heights of its peaks and valleys) at \emph{typical points} on the unit $\ds{N-1}$ sphere in field-space must be $N$-independent. This constraint is inspired by a similar approach in \cite{Greene:2013ida} for the case of quartic potentials. Points on the unit sphere can be parametrized as $\ds{\vec{\phi} = \f{1}{\sqrt{N}}(c_1, c_2, ..., c_N)}$, where each of the $\ds{c_i}$'s are typically $O(1)$\footnote{Points on the unit-sphere also satisfy $\ds{\Sum_{i=1}^N c_i^2 = N}$.}. 

Since the potential is additive, we require each component potential $\ds{V_i(c_i/\sqrt{N})}$ to scale as $1/N$, so that the overall potential (at \emph{typical} points) does not scale with $N$. Therefore, by imposing this particular form of $N$-independence in the potential, we are forced to choose parameters ($u$, $v$, etc.) from distributions such that the $V_i$'s and the length-scales of the $\phi_i$ axes scale with $N$. 

From the arguments in the preceding section, we know that tunneling happens along a field axis (the one which minimizes the action). Therefore, the barrier height and barrier width on the \emph{tunneling trajectory} also scale as $1/N$ and $1/\sqrt{N}$, respectively (due to the fact that a field axis is \emph{not} a typical direction). Here $N$ plays the role of $g$ in our scaling arguments, and we are left with the scaling exponents $\ds{a=-1/2}$ and $\ds{b=-1}$. This automatically fixes $c=0$, i.e., the typical bounce radius is $N$-independent. The exact bounce action (\ref{eq:bounce}) scales as $N^{-1}$, and so does the approximation (\ref{eq:approx}). Thus, for these potentials, tunneling probability is enhanced as $N$ grows. This agrees with the result found in \cite{Greene:2013ida} where the action scales as $N^{-\alpha}$ with $\alpha>1$.

\subsection{Multi-Field Potentials With Cross Couplings}

In the case of additive potentials, the bounce solution satisfies the longitudinal equation of (\ref{eq:eom3}) because it is effectively a single field bounce. The transverse equation is trivially satisfied because the path is a straight line and the variation of the potential in all the remaining $N-1$ directions is also $0$ (all the remaining field coordinates rest at a minimum of their potential)\footnote{Note that for the non-axis solutions to (\ref{eq:eom2}), the LHS and RHS of the second equation are typically nonzero.}. 

When we move away from additive potentials to potentials with cross-couplings, the story becomes considerably more complicated. For such potentials, straight line directions with no transverse gradients do not generally exist which makes it difficult to identify solutions that satisfy both equations (\ref{eq:eom3}). However, if we are able to identify the bounce trajectory, it is possible to compute either the exact action (\ref{eq:bounce}) or its approximation (\ref{eq:approx}), both of which scale the same way. 

Greene et al. \cite{Greene:2013ida} sought to make the multi-field tunneling problem tractable for quartic potentials by calculating estimate (\ref{eq:approx}) in two steps. In one step, they assumed that the potential profile of the trajectory corresponds approximately to a straight-line path and studied the variation of the bubble radius (presumably $r_\Sigma$ or something similar) for quartic potential profiles. They observed that the radius usually took on values within the same order of magnitude for a distribution of sampled potentials and therefore attributed a standardized value of radius for their bounce action estimate. In the second step, they calculated the surface tension integral for the most ``obvious" choice of tunneling trajectory, the one passing through the smallest surface tension barrier. Our arguments indicate that this step is not justified as it may pick the wrong tunneling direction/trajectory.

Using a code improvised from \cite{Wainwright:2011kj}, we computed the exact action for two-field potentials in order to compare with the estimate of Greene et al. \cite{Greene:2013ida}. For the potentials we considered, their estimate was of the same order of magnitude as the exact bounce action. It would be interesting to check the approximation for larger numbers of fields by numerically computing the bounce action to see if the approximation still agrees.

\section{Conclusions}\label{sec:conclusion}

We introduced a new class of potentials with exact analytic bounce solutions corresponding to tunneling, which could be considered a generalization of the Fubini instanton to non-integer powers. Such potentials could possibly have a role to play in the string theory landscape and therefore these exact bounce solutions may prove valuable for further study.

We used scaling arguments to observe that for tunneling potentials with a certain shape of the potential profile, the following hold true: 
\begin{enumerate}
 \item Making the barrier taller ($\ds{a>0}$) and narrower ($\ds{b<0}$) always lowers the bounce action.
 \item Making the barrier taller ($\ds{a>0}$) and broader ($\ds{b>0}$) can still lead to a lower bounce action if the height increases faster than the width to the fourth power ($\ds{a>4b}$).
\end{enumerate}
Furthermore, we recognize that scaling does not account for the considerable diversity in shapes of barriers, which means that the dependence of action (and bounce radius) on various shapes of potential profiles is still an open and rich problem in its own right.

We also observe that the approximation (\ref{eq:approx}), which involves multiplying the ``surface tension" of the bubble by its 3-dimensional surface area, scales the same way as the bounce action does; therefore, its accuracy will be preserved under any transformation that could be described purely in terms of scaling.

Finally, we note that the intuition from single field potentials directly translates to the case of additive multi-field potentials, where the bounce trajectory lies along one of the field axes. For general multi-field potentials, identifying the actual bounce trajectory is still an open problem and we do not yet have a simple way of calculating or estimating the bounce trajectory corresponding to tunneling out of a false vacuum.

\section*{Acknowledgments}
We would like to thank J. Distler, G. Elor, M. Endres and A. Masoumi for helpful discussions. SP thanks the Center for Theoretical Physics at MIT for its warm hospitality while this work was  completed. This material is based upon work supported by the National Science Foundation under Grant Number PHY-1316033, by a grant from the Simons Foundation ($\#$305975 to Sonia Paban), and by the U.S. Department of Energy under grant Contract Number DE-SC00012567.


\end{document}